**György Csomós[1,\*], Balázs Lengyel[2,3]**

[1] University of Debrecen, Ótemető u. 2-4, 4028 Debrecen, Hungary
[2] Agglomeration and Social Networks Research Lab, Centre for Economic and Regional Studies, Hungarian Academy of Sciences, Tóth Kálmán u. 4, 1097 Budapest, Hungary
[3] International Business School Budapest, Záhony u. 7, 1031 Budapest, Hungary
\* Corresponding author: csomos@eng.unideb.hu

**Mapping the efficiency of international scientific collaboration between cities worldwide**

International scientific collaboration, a fundamental phenomenon of science, has been studied from several perspectives for decades (Ponds, 2009; Wagner & Leydesdorff, 2015; Zhang et al., 2018). There is evidence that international scientific collaboration started in the nineteenth century; however, due to the rapid globalization of science, it has gained significance only in recent decades, and its growth rate is still accelerating (Luukkonen et al., 1993; The Royal Society, 2011; Wagner et al., 2017). Analysis of the spatial aspects of the science system, including that of international scientific collaboration, by using bibliometric data is in the scope of spatial scientometrics (Frenken et al., 2009; Gao et al., 2013). Spatial scientometric analysis most often focuses on examining international scientific collaboration between countries and regions, but has shown little interest in examining the city level. One reason for this discrepancy is that cities are considered the most inhomogeneous spatial elements of the science system, where organizations are generally conducting research and producing publications independently from each other. Furthermore, problems stemming from the non-standardized territorial demarcation of cities (a permanent hot topic of urban geography) make the results of spatial scientometric analysis at the city level rather uncertain (for example, the Web of Science, one of the well-known indexing databases, neglects to present bibliometric data on the city level even if it could do).

Despite the aforementioned problems, several spatial scientometric studies focusing on cities have been published in the last two decades. One part of these studies examines the position of cities as nodal points of the science system based on total publication output (quantity approach) (Andersson et al., 2014; Matthiessen & Schwarz, 1999; Csomós, 2018), while another part focuses on mapping cities as centres of excellence in terms of their citation impact (quality approach) (Bornmann & Leydesdorff, 2011; Bornmann & Leydesdorff, 2012). Only a minority of spatial scientometric studies are interested in exploring collaboration patterns of cities, by for example examining internal collaborations (Ma et al., 2014; Chen et al., 2015) or mapping collaboration networks on the basis of co-authored papers (Maisonobe et al., 2016).

In this visualization, we go beyond analyzing cities' international scientific collaboration patterns that are based on the well-known approaches and rather present how efficient the collaborations between cities are. The analysis involves 245 cities in which authors produced at least 10,000 publications during 2014–2016. The name of cities (or that of any spatial units between the institution and the country/state level) can be found in the addresses reported by the authors of publications. We focused on only those city-to-city collaborations that produced at least 300 publications during the aforementioned period of analysis (i.e., an average of 100 publications per year). The collaboration matrix of 245 x 245 cities contains a total number of 7,718 international collaboration links that meet the above criteria. Efficiency of the collaboration between each city-dyad corresponds to the ratio of highly cited papers to all papers produced by co-authors affiliated with those cities between 2014 and 2016. We assume that the higher the efficiency of the collaboration between two cities is, the more likely it is that researchers affiliated with those cities conduct joint research resulting in new scientific breakthroughs.

We group city-to-city links along two dimensions: the 80th percentile is used to partition by number of collaborations and fraction of highly cited papers. The groups contain 4997 links (low collaboration, low efficiency indicated by dark blue in Figure 1), 1180 links (low collaboration, high efficiency indicated by light blue), 1178 links (high collaboration, low efficiency indicated by light red) and 363 links (high collaboration, high efficiency indicated by dark red). We colour the edges by these four groups, set East-West edge curves positive (clockwise) for links with collaborative papers exceeding the 80th percentile and negative (counterclockwise) otherwise and plot the edges on top of each other.

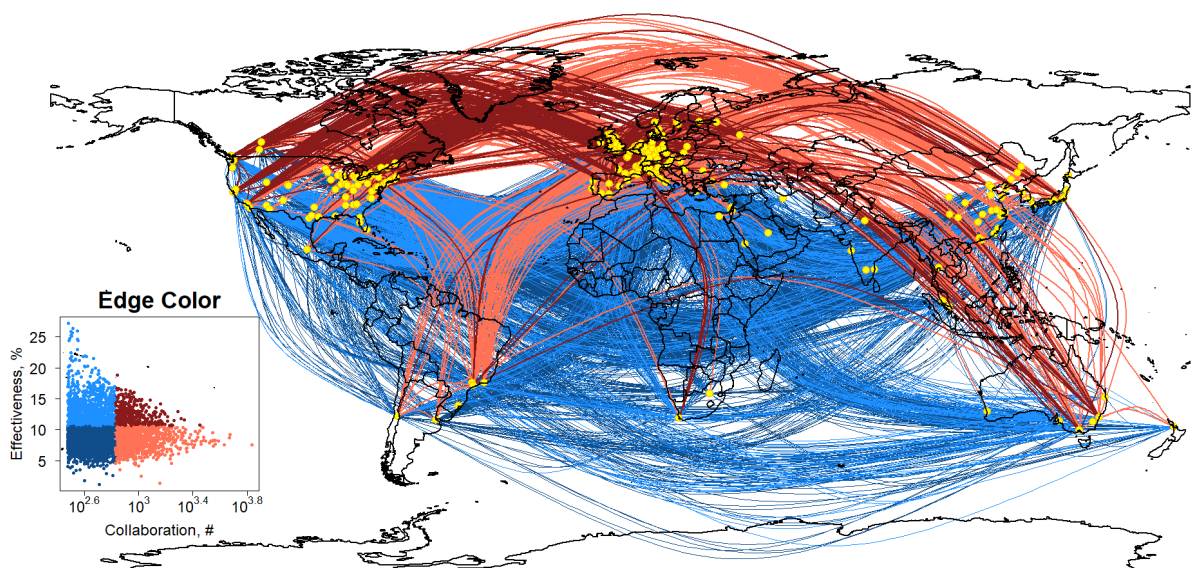

Figure 1. Visualization of efficiency of international scientific collaboration of cities

The strength of the scientific collaboration between two cities in terms of the number of co-authored papers is significantly influenced by the total publication output of organizations located in those cities, the geographical proximity, the historical, cultural, and linguistic ties and the most productive scientific disciplines (Csomós, 2018). However, these factors do not or only partly provide appropriate explanations for the varying efficiency of collaborations between certain city-dyads.

Figure 1 shows that the higher the number of co-authored papers between two cities (more precisely: between authors affiliated with that cities) is, the more likely it is that the efficiency of the collaboration approximates the average value. For example, the London-Paris collaboration link is the highest in the world in terms of the number of co-authored papers (more than 6800 papers produced between 2014 and 2016), but the efficiency of the collaboration (7.617) is around the average value (9.062). The scientific collaboration between Northern American, Western European and Australian cities having high total publication output generally produces a large number of co-authored papers from which many co-authored papers become highly cited, and for this reason the efficiency of these collaborations approximates (or is slightly above) the average value (most of these links are indicated by dark red in Figure 1).

The scientific collaborations between the leading Latin-American cities (e.g., Sao Paulo, Rio de Janeiro and Mexico City), Chines cities (e.g., Beijing, Shanghai and Hong Kong) and other East Asian cities (e.g., Tokyo, Osaka, Seoul and Singapore), but primarily with western cities, have recently been significant, however the efficiency of these collaborations generally remains below average (these links are indicated by light red in Figure 1). These emerging cities (Japanese cities as well) have established high scientific collaboration in terms of the number of co-authored papers with many cities in the world but only few of these collaborations produce new scientific breakthroughs (which is indicated by the low number of highly cited papers).

Collaborations between cities having the highest efficiency are generally based upon a fewer number of (approximately 300–400) co-authored papers, while typical geographical patterns cannot be detected (these links are indicated by light blue in Figure 1). The Toulouse-Copenhagen collaboration produces the highest efficiency (27.213) in the world exceeding the average value by three times. The efficiency of the collaboration between Warsaw and Nijmegen (26.380), Padua and Toronto (25.989) and Helsinki and Montreal (25.868) is also very high. None of the geographical proximity (as for the case of the London-Paris, and the Seattle-Vancouver, BC collaboration links), the historical, cultural and linguistic ties (as for the case of the Copenhagen-Stockholm, and the Paris-Montreal collaboration links), or the size of the output (as in the case of the Beijing-Tokyo, and the London-Boston collaboration links) are factors influencing the efficiency of these collaborations. It is, however, more important to know which are the most productive scientific disciplines in those cities. For example, in each aforementioned case, the most productive discipline is 'Astronomy and Astrophysics' (40–50 percent of all co-authored papers are produced in that field) of which the ratio of highly cited papers to all papers is generally very high. Because in these collaborations the number of co-authored paper is low, the attitude of individual researchers located in those cities becomes even more important (i.e., the collaboration between two or some star researchers becomes more visible).

Most international scientific collaborations of city-dyads produce only a few numbers of co-authored papers (i.e., 300–600 papers) with low efficiency (these links are indicated by dark blue in Figure 1). In these cases, pronounced geographical patterns cannot be found.

**Software**
We used 'R' for both network visualization and mapping.

**Data sources**
All data were obtained from the Web of Science's (owned by Clarivate Analytics) Science Citation Index Expanded, Social Sciences Citation Index and Arts & Humanities Citation Index databases.

**Dataset**
Full dataset on the efficiency of international scientific collaboration of cities is available at Harvard Dataverse (https://doi.org/10.7910/DVN/9E6Y1M).

**Declaration of conflicting interests**
The authors declared no potential conflicts of interest with respect to the research, authorship, and/or publication of this article.

**Funding**
The authors received no financial support for the research, authorship, and/or publication of this article.

**References**

Andersson DE, Gunessee S, Matthiessen CW and Find S (2014) The geography of Chinese science. *Environment and Planning A* **46**(12): 2950−2971.
Bornmann L and Leydesdorff L (2011) Which cities produce more excellent papers than can be expected? A new mapping approach, using Google Maps, based on statistical significance testing. *Journal of the American Society for Information Science and Technology* **62**(10): 1954−1962.

Bornmann L and Leydesdorff L (2012) Which are the best performing regions in information science in terms of highly cited papers? Some improvements of our previous mapping approaches. *Journal of Informetrics* **6**(2): 336−345.

Chen W, Xiu C, Liu W, Liu Z, and Yu Z (2015) Visualizing intercity scientific collaboration networks in China. *Environment and Planning A* **47**(11): 2229−2231.

Csomós G (2018) A spatial scientometric analysis of the publication output of cities worldwide. *Journal of Informetrics* **12**(2): 547−566.

Frenken K, Hardeman S and Hoekman J (2009) Spatial scientometrics: Towards a cumulative research program. *Journal of Informetrics* **3**(3): 222−232.

Gao S, Hu Y, Janowicz K and McKenzie G (2013) A spatiotemporal scientometrics framework for exploring the citation impact of publications and scientists. In: *GIS: Proceedings of the ACM International Symposium on Advances in Geographic Information Systems*, Orlando, FL, 05-08 November 2013, pp. 204−213.

Glänzel W (2001) National characteristics in international scientific co-authorship relations. *Scientometrics* **51**(1): 69−115.

Luukkonen T, Tijssen RJW, Persson O and Sivertsen G (1993) The measurement of international scientific collaboration. *Scientometrics* **28**(1): 15−36.

Ma H, Fang C, Pang B and Li G (2014) The effect of geographical proximity on scientific cooperation among Chinese cities from 1990 to 2010. *PLoS ONE* **9**(11),e111705

Maisonobe M, Eckert D, Grossetti M, Jégou L and Milard B (2016) The world network of scientific collaborations between cities: domestic or international dynamics? *Journal of Informetrics* **10**(4): 1025−1036.

Matthiessen CW and Schwarz AW (1999) Scientific centres in Europe: An analysis of research strength and patterns of specialisation based on bibliometric indicators. *Urban Studies* **36**(3): 453−477.

Ponds R (2009) The limits to internationalization of scientific research collaboration. *Journal of Technology Transfer*, **34**(1): 76−94.

The Royal Society (2011) Knowledge, Networks and Nations: Global scientific collaboration in the 21st century. The Royal Society, London. Available at: https://royalsociety.org/topics-policy/projects/knowledge-networks-nations/report/ (accessed 03 August 2018)

Wagner CS and Leydesdorff L (2005) Network structure, self-organization, and the growth of international collaboration in science. *Research Policy* **34**(10): 1608−1618.

Wagner CS, Whetsell TA and Leydesdorff L (2017) Growth of international collaboration in science: revisiting six specialties. *Scientometrics* **110**(3): 1633−1652.

Zhang C, Bu Y, Ding Y and Xu J (2018) Understanding scientific collaboration: Homophily, transitivity, and preferential attachment. *Journal of the Association for Information Science and Technology* **69**(1): 72−86.